\documentclass[letterpaper,twocolumn,10pt]{article}
\usepackage{usenix-2020-09}

\usepackage{subfiles}
\usepackage{times,url,color,soul,xspace,enumitem,booktabs}
\usepackage{graphicx}
\hypersetup{breaklinks=true}

\newcommand{\namex}{Griffin}

\AtBeginDocument{%
  \providecommand\BibTeX{{%
    \normalfont B\kern-0.5em{\scshape i\kern-0.25em b}\kern-0.8em\TeX}}}

\begin{document}

\title{A Case for a Programmable Edge Storage Middleware}

\author{
{\rm Giulia Frascaria, Animesh Trivedi, and Lin Wang}\\
Vrije Universiteit Amsterdam
}

%
%
%

\maketitle


\begin{abstract}
Edge computing is a fast-growing computing paradigm where data is processed at the local site where it is generated, close to the end-devices. 
This can benefit a set of disruptive applications like autonomous driving, augmented reality, and collaborative machine learning, which produce incredible amounts of data that need to be shared, processed and stored at the edge to meet low latency requirements. However, edge storage poses new challenges due to the scarcity and heterogeneity of edge infrastructures and the diversity of edge applications.
In particular, edge applications may impose conflicting constraints and optimizations that are hard to be reconciled on the limited, hard-to-scale edge resources. 
In this vision paper we argue that a new middleware for constrained edge resources is needed, providing a unified storage service for diverse edge applications. 
We identify programmability as a critical feature that should be leveraged to optimize the resource sharing while delivering the specialization needed for edge applications. 
Following this line, we make a case for eBPF and present the design for \namex{}---a flexible, lightweight programmable edge storage middleware powered by eBPF.
\end{abstract}

\section{Introduction}\label{sec:introduction}

Cloud computing and its associated infrastructure services (for storage and compute) have been the standard for computing for the past two decades. However, in recent years we have witnessed a shift in trends that goes under the name of edge computing~\cite{satyanarayanan2009case}, that pushes computation and storage resources towards the network edge, away from the central clouds.
In edge computing end-users are in close proximity to edge data centers (typically referred to as cloudlets or edge nodes), allowing to process data on the spot rather than transferring it to the remote cloud.
This setup facilitates a set of disruptive applications that benefit from reduced end-to-end latency and network traffic, which can be derived from an efficient edge infrastructure~\cite{satyanarayanan2017emergence}. 
Examples of such applications are collaborative machine learning, autonomous driving, augmented reality and live video analytics~\cite{2017-hotcloud-keynote-satya-edge}.



The centerpiece characteristic of these applications is that various end-devices or edge nodes generate a vast quantity of data that needs to be processed in due time. Overall, storing and processing significant amounts of data with a diverse set of applications is not a unique paradigm to edge computing. 
In the cloud realm we witnessed how the Big Data trend fuelled innovation for a variety of specialized services, such as blob storage with infinite store (e.g., S3~\cite{s3}), elastic key-value stores~\cite{2020-fast-infinicache,amazon-elastic-store}, disaggregated storage~\cite{2017-nsdi-decibel}, stores with multiple APIs (block, key-value, timeseries databases) with multi-consistency storage systems~\cite{lakshman2010cassandra} to cater the needs for different cloud data-processing applications. 
However, the unique deployment scenario of edge computing makes continuing this cloud-centric, multiple-service storage model (different storage services for different needs) a challenging task.


(i) First, unlike large central clouds equipped with abundant resources and specialized hardware~\cite{facebook-opencompute,ibm-openpower}, edge nodes are more likely to be constrained by physical limitations, e.g., a cellular base station may only have space or power supply to host a handful of commodity servers.
Thus, edge nodes are capacity-limited by construction and cannot be scaled infinitely.
As a result, it would be impractical to deploy half a dozen services at the edge to fit different edge application use cases. 
In addition to it being resource-intensive, deploying multiple services also works against their availability. While different cloud storage services typically do not share storage hardware and are deployed in different fault domains for maximum availability~\cite{aws-fault-domains, azure-fault-domains} this would not be feasible with the restricted resources of edge nodes. 
Given the infrastructural constraints at the edge, it would be inefficient to deploy multiple edge storage services tailored for a specific edge applications as it is currently possible in the cloud. 


(ii) Second, a strong trend in the cloud is resource dis-aggregation which decouples storage (among other resources) from the compute~\cite{2016-osdi-disaggregation}. 
However, due to the limited (storage and compute) resources at the edge and the preference for Commodity Off-The-Shelf (COTS) devices to deploy edge nodes, we are witnessing an opposite shift toward ``re-aggregation'' of storage and compute resources, where data should be processed in-place with the least amount of movement. 
This calls for the need for functionalities like secure multi-tenancy, data management, access control, and data ownership maintenance in any edge storage system. 
Moreover, edge applications also expose different data management goals.
For example, online edge gaming applications may prioritize data movement with the client mobility, whereas a medical application may necessitate on-premise data processing with zero data movement off-premise. 
In these examples, data processing is highly coupled with how the data is managed and staged for processing.


(iii) Last, different edge applications have different data access requirements~\cite{leidall2019edge,griffin}. For example, an edge machine learning (ML) application can benefit from having an ML-specific storage API~\cite{2019-socc-cirrus}, whereas for machine vision applications a simple key-value API would be sufficient~\cite{2018-hotedge-machinevision}. Hence, an ideal edge storage service must be able to support a variety of APIs, abstractions, access semantics, and potential application-specific policy optimizations regarding data mobility and lifetime.


In this vision paper, we make a case for a single storage service on shared edge storage resources with application-specific customization for semantics and data management. We argue that the seemingly impossible requirements  of sharing and customization can be attained with a  programmable  storage service. Unlike hardware programmability (e.g., just FGPA~\cite{2019-atc-insider,2014-osdi-willow,2016-isca-biscuit}, or
a mix of FPGAs, GPUs and ASICs at the Edge as presented by Theophilus A. Benson in his EdgeSys'20 keynote~\cite{2020-edgesys-keynote}) in this work we argue for ``software'' programmability using the eBPF language support and runtime~\cite{2020-middleware-storage-nbd,bpftales,datadeluge}. An element of  such language-supported and run-time programmability has been explored for compute management at the edge, e.g., Sledge~\cite{2020-middleware-sledge}. In this work we push such capabilities to the storage and present a design for an \textit{eBPF-powered programmable edge storage middleware}. As we will expand in Section \ref{sec:programmability}, the use of eBPF brings a set of unique features that make it desirable for a programmable storage service for the edge.





\section{Edge Storage: A New Start?}\label{sec:differences}
 \begin{table*}[]
\centering
\small
\begin{tabular}{lll p{8cm} }
\toprule
\textbf{Property} & \textbf{Cloud} & \textbf{Edge} & \textbf{Implications}
\\
\midrule
storage resources & abundant & scarce & can not deploy multiple, separate
storage services \\
deployment model & disaggregated & aggregated & need support for (i) 
secure execution of multi-tenant compute and storage logic, (ii)
less variable QoS and SLA objectives\\ 
mobility  & absent & likely & need support for application-dependent data/state mobility
\\
resource types & mostly homogeneous & mostly heterogenous & need support
for diverse device types and ISAs \\ 
storage API & multiple supported & multiple needed & need support
application-dependent customizations\\
maintenance & provider dependent &   deployment dependent &  need support
deployment-dependent customizations \\ \bottomrule
\end{tabular}
\caption{Properties of cloud and edge storage and their
implications.}
\label{tab:compare}
\end{table*}

The edge is composed of \textit{distributed}, 
\textbf{heterogeneous} nodes that operate with potentially limited resources. In contrast, the cloud has 
abundant resources which enable users to deploy arbitrary 
applications without worrying about resource exhaustion. 
In a way, cloud computing is a ``double-blind'' deployment where 
applications do not need to worry about cloud infrastructure maintenance, 
and cloud providers do not need to worry about application 
scalability due to the resource abundance.
Instead, the scarcity of edge resources forces both applications and edge 
providers to revisit basic operational assumptions and revise design 
implications for that. Table \ref{tab:compare} summarizes our 
discussion in this section.

\textbf{Storage resources:} Cloud data centers have an abundance of storage capacity (10-100 TBs), servers (in thousands), and network bandwidth (40/100/200 Gbps) between servers. 
Consequently, a cloud provider can support multiple storage services to accommodate multiple workloads while assuming virtually unbound storage (and compute) resources. 
In contrast, edge nodes typically have much more modest configurations. For example, Dell Edge gateway 3000 series contains a 2-core Intel Atom processor, with 2 GB DRAM and 64 GB flash storage~\cite{2021-dell-edge-gateways}. 
Furthermore, edge storage resources can not be easily scaled out because of the geographical dispersion and limited network bandwidth. As a result, unlike the current cloud deployments, it is likely that we can not deploy multiple edge storage services such as a key-value store, a shared file system, shared block storage, and a cache, on the edge, simultaneously. A more practical solution would be to build a single service that can support a multitude of use cases in a unified but customizable manner.

\textbf{Deployment model:} Cloud deployments typically use a disaggregated storage deployment model where compute and storage nodes are not co-located, but kept separate and connected through high-speed networks~\cite{2017-hotsotrage-sto-dissg}. This improves the utilization by better storage provisioning. However, due to the limited resources at the edge, we are more likely to see an \emph{aggregated} deployment model where compute and storage functions are co-located. Such a deployment model presents two key challenges. First, due to the multi-tenant nature of edge computing, the storage service must support a secure way in which multiple untrusted functions can co-exist on the edge.
Second, the storage service must ensure that edge functions do not over-consume compute/storage resources, with deterministic quality of service (QoS) and service-level agreements (SLAs) on data processing. 
Delivering these objectives would require constantly monitoring the system status to optimize for smart data placement and function execution to meet application access demands.

\textbf{Mobility:} Cloud data centers are geographically distributed, but they are still designed around a hyper-converged architecture that serves wide geographical areas.
Edge nodes instead are smaller and more distributed in the field, increasing the likelihood of a user moving from one node to another. 
In latency-sensitive applications accessing the closest edge node is a desirable feature.
While cloud data centers do support a limited form of data mobility where they follow usage patterns (e.g., diurnal)~\cite{2018-osdi-akkio}, these efforts are focused on the daily or weekly patterns, rather than seconds or milliseconds which may be needed at the edge. 
As the user moves, the storage availability in the closest node can also vary. 
For this reason, it is relevant to ensure that active storage management can maintain a buffer of storage space for moving users and remove old and unused data from storage. 
Furthermore, a user should be able to express a limited or no mobility preference to preserve privacy, since sensitive data may not be allowed to leave specific edge node premises. 
Nonetheless, the end-device mobility pattern, frequency, data retention goals, and the amount of storage needed is very application-dependent features and require an API to express these features as storage system policies regarding data mobility.

\textbf{Resource types:} Traditionally cloud data centers prefer to host homogeneous machines for the ease of maintenance and economy of scale~\cite{2016-hyperscale-homogeneous-dc1}. 
In contrast, the edge represents a very heterogeneous deployment environment for any service~\cite{2019-csur-rm-edge}, and the notion of heterogeneity goes beyond the storage capacity per node and the speed of the storage technology. 
As we discussed previously, we are more likely to see an aggregated compute-storage deployment of applications at the edge. 
These applications can run on multiple CPUs of different ISAs (x86, ARM, POWER) and accelerators (GPUs, FPGAs, TPUs, or ASICs)~\cite{2020-vee-edge-isa}. 
As a result, a storage system should be able to support low-level device and CPU management, code optimization techniques, execution policies (performance, energy, cost)---many or all of them cannot be hard-coded and pre-optimized without the knowledge of the final deployment hardware resources and architecture. Consequently, techniques from programming languages literature such as JIT compilation and language/runtime supported IR representations would be an important avenue to explore for supporting edge-heterogeneity for storage customization.

\textbf{Storage API:} 
As a direct consequence of having an abundance of networking and storage resources, cloud data centers have specialized services catered for different application needs~\cite{2018-alluxio}. 
Examples of such systems include file systems (Luster, PVFS, GPFS), various blob, object, and key-value stores (Ceph, Swift, Redis, Memcached), NoSQL DBs, etc. 
However, edge storage systems are still in development and in a need of application-specific customization to meet their needs~\cite{griffin}. 
When building a single storage service, we need to support a multitude of application data and state access requirements (discussed in detail in our previous work~\cite{griffin}): multiple APIs (KV, timeseries, transactions, events, and CRDTs) with support for application-specific mobility policies, data-access policies (sharing), etc. 
For example, in the case of concurrent updates to an object, an application can define its customized policy regarding if the first write should win, or the last one. 
Similarly, based on the value of the data (compute data duality) and application resiliency (e.g., ML training), applications can decide how to tolerate faults in case of an edge storage node failure. 
While it is yet to be explored up to what extent a single unified edge storage service can be designed and implemented, we highlight that there is a similar push inside cloud data centers for single unified services~\cite{2019-atc-nodekernel}.

\textbf{Maintenance:} Cloud storage services strive to keep multiple 9s uptimes, and hence invest heavily in the deployment and maintenance of the storage hardware and services. 
Data is backed up and replicated, and the lifetime of storage media within cloud data centers is continuously monitored by the cloud provider who takes care of transparently substituting the worn-out infrastructure while migrating the data. 
Old storage devices are then physically destroyed at the end of their lifecycle to prevent any possibility of data recovery. 
In edge computing landscapes this is an infeasible maintenance commitment, due to the dispersion of the infrastructure. 
It is then crucial that some degree of ``active'' health monitoring is implemented in the infrastructure in order to facilitate maintenance operations and to avoid data loss. 
This is especially relevant since it is not possible to assume active, continuous maintenance of edge nodes as we can assume for cloud storage services.

To summarize, so far we systematically analyze fundamental properties that have changed when moving from the cloud to the edge, and how these changes affect the design of a storage service for the edge.
We are clearly not the first one to make a case for an edge store service~\cite{mortazavi2018pathstore,mayer2017fogstore}. However, prior work only addresses a subset of the edge storage requirements~\cite{griffin}. In the next section we explain how a programmable storage middleware can help to meet the requirements of edge applications.

\section{Programmable Edge Storage}\label{sec:programmability}


Based on the discussion so far, the central premise of this paper is: Can we build a unified storage service for the edge that can be customized to serve all edge applications efficiently? In this section we argue it is possible, since by leveraging \textit{programmability} of a storage service as its first class citizen the storage middleware can allow users to customize to their needs. 
In particular, we make a case for \emph{software-based programmability} and show how the extended Berkeley Packet Filter (eBPF) can be leveraged to address the edge storage challenges.
We then present the design of \namex{}---our proposed edge storage middleware powered by eBPF.

\subsection{Why eBPF for Edge Programmability}
eBPF, short for extended Berkeley Packet Filter~\cite{mccanne1993bsd}, is a software infrastructure that operates within the Linux kernel. BPF has been used for user-defined packet filtering for decades, but was extended and redesigned for additional functionalities. The extended BPF enables use cases such as advanced network packet processing, monitoring, tracing, and security~\cite{cilium, katran, falco}, and is 
now being expanded constantly in the Linux kernel development efforts.
We argue that the versatility of eBPF makes it a good candidate for designing a programmable edge store due to:  
\begin{itemize}[leftmargin=10pt,itemindent=0em,nolistsep]
  \item\textit{Expressiveness:} Previous work on eBPF applicability in the storage domain highlights its performance and flexibility benefits~\cite{barbalace2019extos,bijlani2019extension}. The basic instruction set is expressive to capture many common storage-related usecases like aggregation, filtering, transformation, etc. These functions use cases can be attached to the execution of virtually any function within the Linux kernel 
  and can be (re)programmed without halting the system.
  \item\textit{Wide availability:} The eBPF toolchain is supported by the Linux kernel, and the clang and gcc compilers. Hence, it is immediately available on any device supporting Linux (including sensors and IoT). This availability decreases the inherent inertia when deploying a new software stack. 
  \item \textit{Secure and bounded multi-tenant execution:} Thanks to its  simple(r) ISA, eBPF is amenable for verification and extensions~\cite{2019-pldi-ebpf}. The current Linux/eBPF toolchain can inject and run user-defined code in the 
  Linux kernel in a safe way by providing symbolic execution and termination guarantees that ensure the extension is safe and will not stall in lengthy or infinite computation. Reusing the rest of the Linux's isolation machinery, we can ensure a safe and secure multi-tenancy execution of storage customization logic for all applications.  
  \item \textit{A unified ISA for all:} Currently eBPF toolchains exist for multi-arch CPUs, (smart) NICs and switches (P4 supports eBPF compilation), and even FPGAs~\cite{2020-osdi-hxdp,2020-csur-ebpf-xdp} with a
  support for JITing. As a result, we believe that eBPF is what comes the closet to a unified ISA with support for heterogenous computing and I/O devices. Such unified support also open possibilities for a unified optimizations across the network-storage stacks. For example, network support can be use to replicate data packet necessary for storage replication in a distributed setting.
\end{itemize}

\noindent\textbf{What are the alternatives:} Alternatives would have been hardware-supported programmability (FPGAs, ISCs, ASICs)~\cite{2019-atc-insider,barbalacecomputational}, though their broader applicability is yet to be seen at the edge~\cite{2018-hotedge-fpga}. We also look into other language provided isolation like with Rust, Java script, or WebAssembly. Such techniques have been used in the context of storage~\cite{2018-osdi-splinter,2020-socc-sfunct} in data centers, and with compute on the edge as well~\cite{2020-middleware-sledge}.
However, they (i) have high runtime overheads (JVM); (ii) does not support enhancing kernel storage routines; (iii) do not support multiple devices. A complete userspace based solution with a lightweight virtualization support~\cite{2020-nsdi-firecracker} would be possible, however, building such infrastructure depends a lot on underlying hardware capabilities (support for hardware virtualization) of edge node devices. However, in a modular system, the mechanism for programmability can be changed to a better alternative, if necessary.

\begin{figure}
  \centering
  \includegraphics[width=\columnwidth]{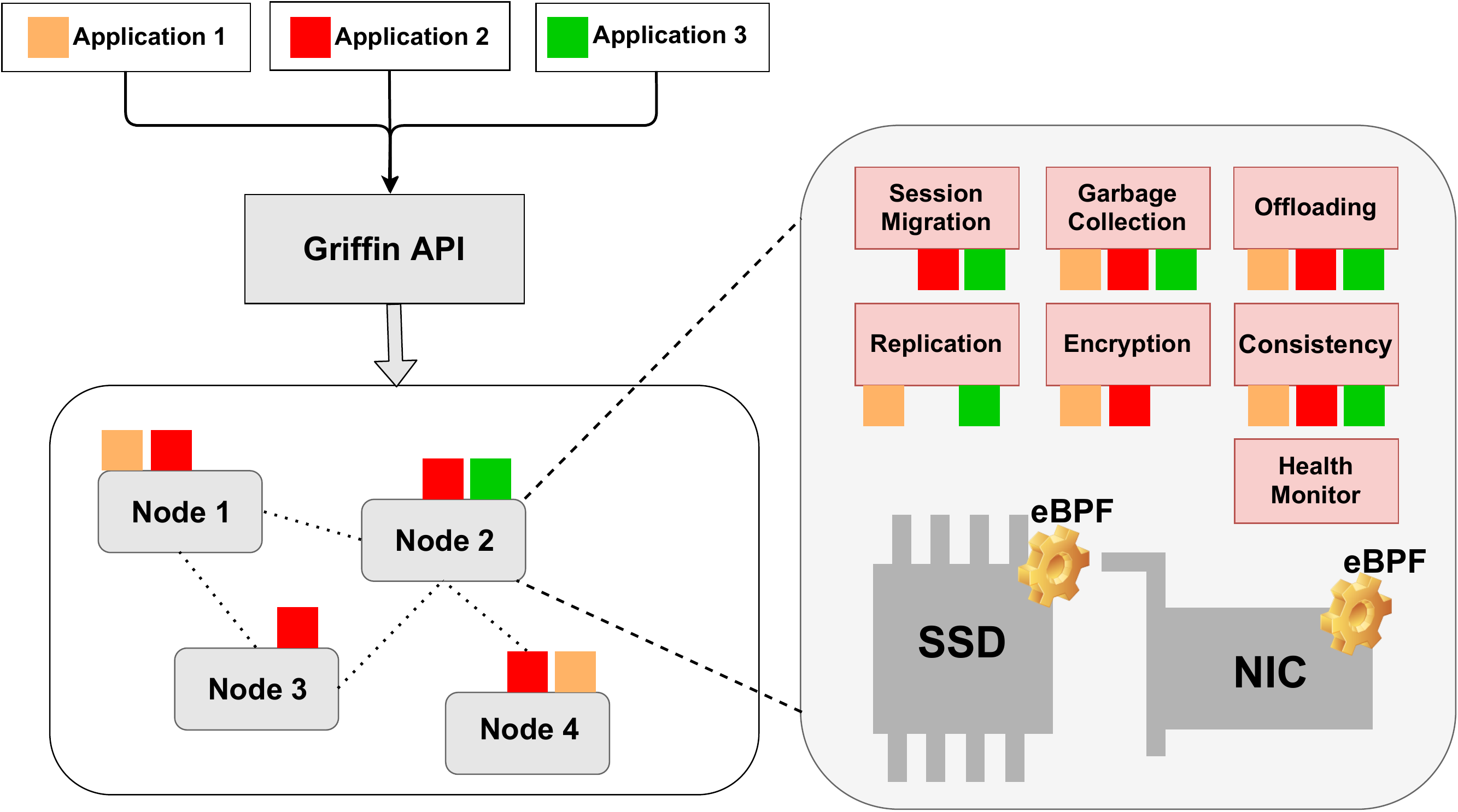}
  \caption{An overview of the Griffin design.}~\label{fig:griffin}
\end{figure}

\begin{table*}[]
\centering 
\small 
\begin{tabular}{ll}
\toprule
\textbf{API} & \textbf{Description} \\ \midrule
\textit{ret = execute(ac, object)} & execution of a application-defined ac 
logic on an object, the basic eBPF functionality \\
\textit{t = register(ac\_trigger, event | code\_path)} & a generic trigger
registration on certain event or code path (similar to the eBPF) \\  \textit{replica\_list = replica\_ac(nodes, state)} & selection of replica
nodes based on any arbitray user criteria \\ 
\textit{new\_replica\_list = loadbalancer\_ac(replica\_nodes, state)} &
ac executed with the current replicas, state, and outputs the new list \\
\textit{\{state, action\} = consistency(object, new\_data, state)} & ac 
takes an object, new data, and the current state, and returns a new
state and action  \\ 
\textit{new\_replica\_list = migration\_ac(replica\_nodes, t, state)} &
ac take the trigger, old replia set, monitoring state and outputs the new
replica list \\ \bottomrule
\end{tabular}
\caption{\label{tab:API} Abridged \namex{} API. \textit{ac} stands for
eBPF powered AppCode (ac).}
\end{table*}

\subsection{\namex{} Design}


We now present the design of Griffin---our proposed edge storage middleware. 
An overview of the system architecture is depicted in Figure~\ref{fig:griffin}.
\namex{} spans a large set of potentially heterogeneous storage nodes at 
the edge. Our proposed approach with software programmability is very similar to 
the spirit of Malacology~\cite{2017-eurosys-malacology}. Malacology, presents a customized re-use of battle-hardened code of a mature 
and stable codebase of the Ceph file system. Here, we argue for building 
the re-usable pieces using the eBPF language support, which can be
selectively used by applications based on their needs. The code snippets 
can be made available to wider egde application as a library.  

The goal of \namex{} is then to provide an expressive APIs for the user to 
specify their application needs such as data format (e.g., key-value, 
timeseries, graph-based), data lifetime, replication, consistency 
(e.g., strong, read-after-write, or eventual), and common or customized 
data operations with service-level objectives (e.g., latency).
In the following, we explain the system-wide services, such as
data replication, consistency, and session migration, provided by \namex{}.
We also show how eBPF serves as the fundamental technology to enable 
light-weight programmable storage functionalities including garbage 
collection, encryption, data erasure and replication, and customized 
computation offloading, which are essential in implementing the APIs 
in \namex{}. We refer to these eBPF provided logic as \textit{appcode} 
that the system will run. See Table~\ref{tab:API} for a brief overview of
the API.

\textbf{Computation offloading:} \namex{} introduces a computation offloading service for applications to run customized \textit{computation} appcode directly on the storage device that holds the data needed for the computation. This can benefit a wide range of data-intensive edge applications where moving the data is typically more expensive than performing the computation itself. \namex{} employs eBPF to implement such a service since it allows 
to perform computation in the Linux kernel and supports runtime updates. Despite the limits on the function complexity, eBPF allows to chain user-defined functionalities in order to implement an expressive set  of functionalities that can be offloaded in the kernel. Thanks to this in-kernel function execution it is possible to remove layers of abstraction from the computation. This would help to shrink the overhead of the execution, reducing the latency perceived by the application and improves the CPU utilization for useful computation.

\textbf{Monitoring:}
Apart from the basic code execution, light-weight monitoring of the
infrastructure is the most significant operation that eBPF supports out of the box.  
The monitoring data is very critical to gather and maintain in a light-weight 
manner, and is used in the other storage system service customizations.
To this end, \namex{} provides a health monitoring service to continuous 
monitor the status of the edge storage devices including the CPU utilization, 
storage utilization and health, and network statistics. Furthermore, eBPF/XDP 
can also be used to monitor the network latencies. Based on the collected 
data, the system has some predefined triggers. For example, when a 
storage device is reaching the end of its lifetime, the health monitoring 
service carries out two tasks: 
(i) It notifies the data replication service to find a new device to replicate the data stored on this device and performs complete data erasure. 
(ii) It signals the edge provider about this situation so that the edge provider can perform maintenance in due time.
Observability and monitoring are currently among the most widespread use cases 
of eBPF. We plan to leverage eBPF to implement a health monitoring service 
for the storage middleware so that data safety is always guaranteed and 
timely maintenance can be performed by edge provider.

\textbf{Load balancing and replication:}
Replication of data is a crucial factor to consider in edge computing. 
First, data replication is key to the availability of edge applications.
Edge nodes can become unavailable due to both unfavorable network conditions or system failures. 
With data replication we can ensure an edge application can always have its required data at its disposal and remain operative.
Second, data replication can be used to improve edge application performance.
As we previously highlighted, many edge applications are collaborative and rely on data sharing to achieve their goals, and since users are distributed on the territory placing data on multiple edge nodes can be beneficial to end-to-end latency improvement for as many users as possible. 
On the other hand, the scarcity of edge resources highlights the need to perform replication only when needed, in order to save storage space across edge nodes. 
For this reason we believe that it is necessary to let the users specify the replication policies and preferences, in order to allow the storage middleware to allocate resources in optimally. 
\namex{} features a data replication service which provides an eBPF-powered API to help customize many decisions when doing replication and load balancing. The default policy is not to replicate data. Applications can provide eBPF appcode to enable replication. The input to the code is a set of edge nodes (with further properties like location, load, capacity), and the eBPF
replication code returns a list of nodes where replicas should be placed. The application is free to apply any criteria it sees fit like best locality, the least loaded, or maximum capacity, etc. The head of the list is the primary node,  others are replica in the chain-replication protocol~\cite{2004-osdi-chain} (the only one supported). Similarly, load-balancing appcode can be attached to 
a load balancing trigger (e.g., CPU or network utilization). The code takes the list of replica machines and can output a new replica where the data should replicated. The collection of tigger metrics can be made light-weight and distributed using network-wide eBPF distribution.

\textbf{Consistency:}
\namex{} also builds a consistency management service which is directly attached to the replication service. 
It allows each edge application to specify its required consistency model (e.g., strong, read-my-write, and eventual) via a high-level API and automatically enforces the selected consistency model when data is replicated by the data replication service. eBPF consistency appcode can be attached and executed everytime a read and/or write is issued. The code can be attached for a particular object or objects satisfying certain criteria (name, creation time, or location). The appcode initializes a state associated with the object at the start. Upon execution, the appcode is expected to return the new state associated and action (hold, reject, accept) with the 
object. Using this basic mechanism one can implement multiple consistency models. For example, a state can be associated with a timestamp which can be used to resolve if the write or read should be admitted or rejected from the system. A hold action can be used to wait for a quorum response. Presence or absence of the state can be use to implement the first-writer-wins or the last-writer-wins consistency model. Read-my-write consistency model can be used by comparing the timestamps (vector clocks can also be stored as the
state). Once a read or write is accepted in the system, it will follow the chain replication protocol for data reading and writing. This design is very similar to the client-driven semantic reconciliation of vector-clocked objects in Dynamo~\cite{2007-sosp-dynamo}.



\textbf{Session migration:} 
\namex{} includes a session migration service to handle user mobility. 
The decisions what to migrate and when is taken with the input data from the monitoring service with user-defined triggers. These triggers can be on the capacity, geo-area bounds, load, and any other monitored property of the system. By default, there is no user session migration with the user mobility. A user can register their triggers to represent interests in system properties. When the session migration trigger executes, it takes input the current replica servers, cause of the trigger, and should output a new list of replica servers. \namex{} ensures that happens in a safe and atomic manner (at an appropriate time) by coordinating user traffic between different edge storage nodes.

\textbf{Data erasure and garbage collection:}
While data storage and sharing is fundamental in stateful edge applications, the data may only be useful for the application within a certain time frame. In central clouds it is not crucial to dispose old data immediately since the resources are elastic and virtually infinite (for the user). However, edge computing does not possess such a luxury of resource abundance. Thus, it is important to proactively retrieve storage space as soon as old data is no longer actively being used. Valid options include performing a data backup to a central cloud or simply erasing the data. \namex{} has a dedicated data erasure and garbage collection (GC) service, expanding what is already an essential component of modern storage technologies. This service exposes APIs for the user to explicitly specify the lifetime of the data and the policy to use for GC in their applications (for example once read/write, delete after certain time period, or if certain event has happened). Such lifecycle-based data management is also explored in data center for short-lived data, e.g.,  serverless~\cite{2018-osdi-pocket}. At runtime, \namex{} reclaims 
storage space according to the specified data lifetime and policy defined in eBPF triggers. eBPF can also be used in this case to enforce different GC policy implementations as well as data erasure instrumenting the storage accesses with garbage collection and data erasure behavior at runtime, depending the lifetime of the data.

\section{Conclusion}
Edge computing is emerging as a fundamentally new way how we design, build, 
and deploy our distributed applications. It shifts the center of computing 
from data centers to the data sources and users.  
As a result we have to re-evaluate many basic assumptions made with the design 
of foundational infrastructure services in data centers and must re-design services for the edge. 
In this vision paper we presented our vision for one such service: an eBPF-based programmable storage middleware for edge computing. 
Based on the differences between cloud and edge storage characteristics, 
we propose the design of Griffin as a programmable storage service 
that allows applications to customize policies like replication, consistency, 
garbage collection, as well as to offload part of the computation 
to the storage service in order to improve the latency. We are currently 
in the process of building \namex.



\Urlmuskip=0mu plus 1mu\relax

\bibliographystyle{plain}
\bibliography{sample}

\begin{thebibliography}{10}

\bibitem{katran}
{Open-sourcing Katran, a scalable network load balancer}, 2018.
\newblock
  \url{https://engineering.fb.com/2018/05/22/open-source/open-sourcing-katran-a-scalable-network-load-balancer/}.

\bibitem{aws-fault-domains}
{Increase availability for Amazon Elasticsearch Service by deploying in three
  Availability Zones}.
\newblock
  \url{https://aws.amazon.com/blogs/database/increase-availability-for-amazon-elasticsearch-service-by-deploying-in-three-availability-zones-2/},
  2020.
\newblock Accessed: 2021-2-20.

\bibitem{azure-fault-domains}
{Manage the availability of Linux virtual machines}.
\newblock
  \url{https://docs.microsoft.com/en-us/azure/virtual-machines/manage-availability},
  2020.
\newblock Accessed: 2021-2-20.

\bibitem{ibm-openpower}
{The OpenPOWER Foundation}, 2020.
\newblock \url{https://openpowerfoundation.org}.

\bibitem{amazon-elastic-store}
{Amazon Elastic Block Store}.
\newblock \url{https://aws.amazon.com/ebs/}, 2021.
\newblock Accessed: 2021-2-20.

\bibitem{cilium}
{Cilium: eBPF-based Networking, Observability, and Security}, 2021.
\newblock \url{https://cilium.io/}.

\bibitem{falco}
{The Falco Project: Cloud-Native runtime security}, 2021.
\newblock \url{https://falco.org/}.

\bibitem{facebook-opencompute}
{The Open Compute Project}, 2021.
\newblock \url{https://www.opencompute.org}.

\bibitem{s3}
{What is Amazon S3?}
\newblock
  \url{https://docs.aws.amazon.com/AmazonS3/latest/userguide/Welcome.html},
  2021.
\newblock Accessed: 2021-2-20.

\bibitem{2020-nsdi-firecracker}
Alexandru Agache, Marc Brooker, Alexandra Iordache, Anthony Liguori, Rolf
  Neugebauer, Phil Piwonka, and Diana-Maria Popa.
\newblock Firecracker: Lightweight virtualization for serverless applications.
\newblock In {\em NSDI}, pages 419--434, Santa Clara, CA, February 2020.
  {USENIX} Association.

\bibitem{2018-osdi-akkio}
Muthukaruppan Annamalai, Kaushik Ravichandran, Harish Srinivas, Igor Zinkovsky,
  Luning Pan, Tony Savor, David Nagle, and Michael Stumm.
\newblock Sharding the shards: Managing datastore locality at scale with akkio.
\newblock In {\em OSDI}, page 445–460, USA, 2018.

\bibitem{2020-middleware-storage-nbd}
Antonio Barbalace, Martin Decky, Javier Picorel, and Pramod Bhatotia.
\newblock Blockndp: Block-storage near data processing.
\newblock In {\em Middleware}, page 8–15, New York, NY, USA, 2020.

\bibitem{barbalacecomputational}
Antonio Barbalace and Jaeyoung Do.
\newblock Computational storage: Where are we today?
\newblock In {\em CIDR}, 2020.

\bibitem{2020-vee-edge-isa}
Antonio Barbalace, Mohamed~L. Karaoui, Wei Wang, Tong Xing, Pierre Olivier, and
  Binoy Ravindran.
\newblock Edge computing: The case for heterogeneous-isa container migration.
\newblock In {\em VEE}, page 73–87, New York, NY, USA, 2020.

\bibitem{barbalace2019extos}
Antonio Barbalace, Javier Picorel, and Pramod Bhatotia.
\newblock Extos: Data-centric extensible os.
\newblock In {\em APSys}, pages 31--39, 2019.

\bibitem{2020-edgesys-keynote}
Theophilus~A. Benson.
\newblock {Life on the Edge: Challenges in Specializing and Accelerating the
  Edge}.
\newblock \url{http://cs.brown.edu/~tab/papers/TABenson_EdgeSyS20_KeyNote.pdf},
  2020.
\newblock Accessed: 2021-02-24.

\bibitem{bijlani2019extension}
Ashish Bijlani and Umakishore Ramachandran.
\newblock Extension framework for file systems in user space.
\newblock In {\em ATC}, pages 121--134, 2019.

\bibitem{2018-hotedge-fpga}
Saman Biookaghazadeh, Ming Zhao, and Fengbo Ren.
\newblock Are fpgas suitable for edge computing?
\newblock In {\em HotEdge}, Boston, MA, July 2018.

\bibitem{2020-osdi-hxdp}
Marco~Spaziani Brunella, Giacomo Belocchi, Marco Bonola, Salvatore Pontarelli,
  Giuseppe Siracusano, Giuseppe Bianchi, Aniello Cammarano, Alessandro Palumbo,
  Luca Petrucci, and Roberto Bifulco.
\newblock hxdp: Efficient software packet processing on {FPGA} nics.
\newblock In {\em OSDI}, pages 973--990, November 2020.

\bibitem{2019-socc-cirrus}
Joao Carreira, Pedro Fonseca, Alexey Tumanov, Andrew Zhang, and Randy~H. Katz.
\newblock Cirrus: a serverless framework for end-to-end {ML} workflows.
\newblock In {\em SoCC}, pages 13--24, 2019.

\bibitem{2007-sosp-dynamo}
Giuseppe DeCandia, Deniz Hastorun, Madan Jampani, Gunavardhan Kakulapati,
  Avinash Lakshman, Alex Pilchin, Swaminathan Sivasubramanian, Peter Vosshall,
  and Werner Vogels.
\newblock Dynamo: Amazon's highly available key-value store.
\newblock In {\em SOSP}, page 205–220, New York, NY, USA, 2007.

\bibitem{2021-dell-edge-gateways}
Dell.
\newblock {Edge Gateway 3003} \mbox{Specification}.
\newblock
  \url{https://i.dell.com/sites/doccontent/shared-content/data-sheets/en/Documents/Dell_Edge_Gateway_3000_Series_spec_sheet.pdf},
  2021.
\newblock Accessed: 2021-02-15.

\bibitem{bpftales}
Giulia Frascaria.
\newblock Bpf tales, or why did i recompile the kernel to average some
  numbers?, 2020.

\bibitem{datadeluge}
Giulia Frascaria.
\newblock Can ebpf save us from the data deluge? a case for file filtering in
  ebpf, 2020.

\bibitem{2020-middleware-sledge}
Phani~Kishore Gadepalli, Sean McBride, Gregor Peach, Ludmila Cherkasova, and
  Gabriel Parmer.
\newblock Sledge: a serverless-first, light-weight wasm runtime for the edge.
\newblock In {\em Middleware}, pages 265--279, 2020.

\bibitem{2016-osdi-disaggregation}
Peter~Xiang Gao, Akshay Narayan, Sagar Karandikar, Joao Carreira, Sangjin Han,
  Rachit Agarwal, Sylvia Ratnasamy, and Scott Shenker.
\newblock Network requirements for resource disaggregation.
\newblock In {\em OSDI}, pages 249--264, 2016.

\bibitem{2019-pldi-ebpf}
Elazar Gershuni, Nadav Amit, Arie Gurfinkel, Nina Narodytska, Jorge~A. Navas,
  Noam Rinetzky, Leonid Ryzhyk, and Mooly Sagiv.
\newblock Simple and precise static analysis of untrusted linux kernel
  extensions.
\newblock In {\em PLDI}, page 1069–1084, New York, NY, USA, 2019.

\bibitem{2016-isca-biscuit}
Boncheol Gu, Andre~S. Yoon, Duck{-}Ho Bae, Insoon Jo, Jinyoung Lee, Jonghyun
  Yoon, Jeong{-}Uk Kang, Moonsang Kwon, Chanho Yoon, Sangyeun Cho, Jaeheon
  Jeong, and Duckhyun Chang.
\newblock Biscuit: {A} framework for near-data processing of big data
  workloads.
\newblock In {\em ISCA}, pages 153--165, 2016.

\bibitem{2019-csur-rm-edge}
Cheol-Ho Hong and Blesson Varghese.
\newblock Resource management in fog/edge computing: A survey on architectures,
  infrastructure, and algorithms.
\newblock {\em ACM Comput. Surv.}, 52(5), September 2019.

\bibitem{2018-osdi-pocket}
Ana Klimovic, Yawen Wang, Patrick Stuedi, Animesh Trivedi, Jonas Pfefferle, and
  Christos Kozyrakis.
\newblock Pocket: Elastic ephemeral storage for serverless analytics.
\newblock In {\em OSDI}, pages 427--444, 2018.

\bibitem{2018-osdi-splinter}
Chinmay Kulkarni, Sara Moore, Mazhar Naqvi, Tian Zhang, Robert Ricci, and Ryan
  Stutsman.
\newblock Splinter: Bare-metal extensions for multi-tenant low-latency storage.
\newblock In {\em OSDI}, page 627–643, USA, 2018.

\bibitem{lakshman2010cassandra}
Avinash Lakshman and Prashant Malik.
\newblock Cassandra: a decentralized structured storage system.
\newblock {\em ACM SIGOPS OSR}, 44(2):35--40, 2010.

\bibitem{2017-hotsotrage-sto-dissg}
Sergey Legtchenko, Hugh Williams, Kaveh Razavi, Austin Donnelly, Richard Black,
  Andrew Douglas, Nathana\"{e}l Cheriere, Daniel Fryer, Kai Mast, Angela~Demke
  Brown, Ana Klimovic, Andy Slowey, and Antony Rowstron.
\newblock Understanding rack-scale disaggregated storage.
\newblock In {\em HotStorage}, page~2, USA, 2017.

\bibitem{leidall2019edge}
Zach Leidall, Abhishek Chandra, and Jon Weissman.
\newblock An edge-based framework for cooperation in internet of things
  applications.
\newblock In {\em HotEdge}, 2019.

\bibitem{2018-alluxio}
Haoyuan Li.
\newblock {\em Alluxio: A Virtual Distributed File System}.
\newblock PhD thesis, EECS Department, University of California, Berkeley, May
  2018.

\bibitem{mayer2017fogstore}
Ruben Mayer, Harshit Gupta, Enrique Saurez, and Umakishore Ramachandran.
\newblock Fogstore: Toward a distributed data store for fog computing.
\newblock In {\em 2017 IEEE Fog World Congress (FWC)}, pages 1--6. IEEE, 2017.

\bibitem{mccanne1993bsd}
Steven McCanne and Van Jacobson.
\newblock The bsd packet filter: A new architecture for user-level packet
  capture.
\newblock In {\em USENIX Winter}, volume~46, 1993.

\bibitem{2016-hyperscale-homogeneous-dc1}
Andrew~W. Moore.
\newblock Technical perspective: Jupiter rising.
\newblock {\em Commun. ACM}, 59(9):87, August 2016.

\bibitem{mortazavi2018pathstore}
Seyed~Hossein Mortazavi, Bharath Balasubramanian, Eyal de~Lara, and
  Shankaranarayanan~Puzhavakath Narayanan.
\newblock Pathstore, a data storage layer for the edge.
\newblock In {\em MobiSys}, pages 519--519, 2018.

\bibitem{2017-nsdi-decibel}
Mihir Nanavati, Jake Wires, and Andrew Warfield.
\newblock Decibel: Isolation and sharing in disaggregated rack-scale storage.
\newblock In {\em NSDI}, pages 17--33, 2017.

\bibitem{2018-hotedge-machinevision}
Arun Ravindran and Anjus George.
\newblock An edge datastore architecture for latency-critical distributed
  machine vision applications.
\newblock In {\em HotEdge}, 2018.

\bibitem{2019-atc-insider}
Zhenyuan Ruan, Tong He, and Jason Cong.
\newblock {INSIDER:} designing in-storage computing system for emerging
  high-performance drive.
\newblock In {\em ATC}, pages 379--394, 2019.

\bibitem{2017-hotcloud-keynote-satya-edge}
Mahadev Satyanarayanan.
\newblock Edge computing: Vision and challenges.
\newblock In {\em Keynote talk at HotCloud/HotStorage 2017}, Santa Clara, CA,
  July 2017.

\bibitem{satyanarayanan2017emergence}
Mahadev Satyanarayanan.
\newblock The emergence of edge computing.
\newblock {\em Computer}, 50(1):30--39, 2017.

\bibitem{satyanarayanan2009case}
Mahadev Satyanarayanan, Paramvir Bahl, Ram{\'o}n Caceres, and Nigel Davies.
\newblock The case for vm-based cloudlets in mobile computing.
\newblock {\em IEEE Pervasive Computing}, 8(4):14--23, 2009.

\bibitem{2014-osdi-willow}
Sudharsan Seshadri, Mark Gahagan, Meenakshi~Sundaram Bhaskaran, Trevor Bunker,
  Arup De, Yanqin Jin, Yang Liu, and Steven Swanson.
\newblock Willow: {A} user-programmable {SSD}.
\newblock In {\em OSDI}, pages 67--80, 2014.

\bibitem{2017-eurosys-malacology}
Michael~A. Sevilla, Noah Watkins, Ivo Jimenez, Peter Alvaro, Shel Finkelstein,
  Jeff LeFevre, and Carlos Maltzahn.
\newblock Malacology: A programmable storage system.
\newblock In {\em EuroSys}, page 175–190, New York, NY, USA, 2017.

\bibitem{2019-atc-nodekernel}
Patrick Stuedi, Animesh Trivedi, Jonas Pfefferle, Ana Klimovic, Adrian
  Schuepbach, and Bernard Metzler.
\newblock Unification of temporary storage in the nodekernel architecture.
\newblock In {\em ATC}, page 767–781, USA, 2019.

\bibitem{griffin}
Animesh Trivedi, Lin Wang, Henri Bal, and Alexandru Iosup.
\newblock Sharing and caring of data at the edge.
\newblock In {\em HotEdge}, June 2020.

\bibitem{2004-osdi-chain}
Robbert van Renesse and Fred~B. Schneider.
\newblock Chain replication for supporting high throughput and availability.
\newblock In {\em OSDI}, page~7, USA, 2004.

\bibitem{2020-csur-ebpf-xdp}
Marcos A.~M. Vieira, Matheus~S. Castanho, Racyus D.~G. Pac\'{\i}fico, Elerson
  R.~S. Santos, Eduardo P. M.~C\^{a}mara J\'{u}nior, and Luiz F.~M. Vieira.
\newblock Fast packet processing with ebpf and xdp: Concepts, code, challenges,
  and applications.
\newblock {\em ACM Comput. Surv.}, 53(1), February 2020.

\bibitem{2020-fast-infinicache}
Ao~Wang, Jingyuan Zhang, Xiaolong Ma, Ali Anwar, Lukas Rupprecht, Dimitrios
  Skourtis, Vasily Tarasov, Feng Yan, and Yue Cheng.
\newblock Infinicache: Exploiting ephemeral serverless functions to build a
  cost-effective memory cache.
\newblock In {\em {FAST}}, pages 267--281, 2020.

\bibitem{2020-socc-sfunct}
Tian Zhang, Dong Xie, Feifei Li, and Ryan Stutsman.
\newblock Narrowing the gap between serverless and its state with storage
  functions.
\newblock In {\em SoCC}, page 1–12, New York, NY, USA, 2019.

\end{thebibliography}

\end{document}